\documentclass[runningheads,a4paper]{llncs}

\usepackage{array}
\newcolumntype{L}[1]{>{\raggedright\let\newline\\\arraybackslash\hspace{0pt}}p{#1}}
\newcolumntype{C}[1]{>{\centering\let\newline\\\arraybackslash\hspace{0pt}}p{#1}}
\newcolumntype{R}[1]{>{\raggedleft\let\newline\\\arraybackslash\hspace{0pt}}p{#1}}

\newcommand{\keywords}[1]{\par\addvspace\baselineskip
\noindent\keywordname\enspace\ignorespaces#1}

\begin{document}

\title{Educating Programmers: A Reflection on Barriers to Deliberate Practice}

\titlerunning{Educating Programmers: A Reflection on Barriers to Deliberate Practice}

\author{Michael James Scott and Gheorghita Ghinea}
\institute{Brunel University, London, UK \\  \email{michael.scott@brunel.ac.uk}}

\authorrunning{Michael James Scott \& Gheorghita Ghinea}

\maketitle

\begin{abstract}
Programming is a craft which often demands that learners engage in a significantly high level of individual practice and experimentation in order to acquire basic competencies. However, practice behaviours can be undermined during the early stages of instruction. This is often the result of seemingly trivial misconceptions that, when left unchecked, create cognitive-affective barriers. These interact with learners' self-beliefs, potentially inducing affective states that inhibit practice. This paper questions how to design a learning environment that can address this issue. It is proposed that analytical and adaptable approaches, which could include soft scaffolding, ongoing detailed informative feedback and a focus on self-enhancement alongside skill development, can help overcome such barriers.

\end{abstract}

\section{Introduction}

Recently, there has been a drive to revitalise computing education \cite{g12}, in part, due to criticisms published by The Nesta Trust \cite{lh11} and The Royal Society \cite{f12}. Unfortunately, few beginners find writing code easy and enjoyable \cite{j01,j02}, so crafting an effective learning environment is not a trivial task. Moreover, despite considerable research into programming instruction since the inception of Computer Science as an academic discipline, many learners have not acquired the desired level of competency \cite{sbe83,m+01,tg11}. Even some whom appear to perform well in early tutorials choose not to pursue the discipline \cite{bm05,c06}. Such issues are so pervasive that the British Computer Society (BCS) declared programming a grand challenge for education research \cite{m+05}.

An aspect of this challenge that the authors have encountered is getting learners to engage in frequent practice. Evidence suggests that levels of effort \cite{v05}, comfort \cite{ws01,v05} and depth \cite{s+06} predict success in a first programming course. This is in line with the theory that it can take approximately ten years of deliberate practice to become an expert \cite{ekt93,w96,e06}. Unfortunately, learners often claim that they lack time or have no motivation to do so \cite{km06}. So if deliberate practice is a key element in the acquisition of programming competencies, how do educators create learning environments that successfully encourage practice?

\section{Cognitive-Affective Barriers and Deliberate Practice}

In order to appreciate how to facilitate frequent practice, the barriers that prevent it should be explored. Programming is markedly distinct from other disciplines because proficiency in other areas does not predict success \cite{bl01,eak08} and some believe that there are no effective aptitude tests \cite{m+05,cbl07}, assuming that aptitudes for programming even exist \cite{ekt93,j02}. This is because the learning material sometimes demands something very novel to new learners \cite{h04}, drawing on skills that, at present, are seldom developed prior to programming instruction:

By means of metaphors and analogies we try to link the new to the old, the novel to the familiar. Under sufficiently slow and gradual change, it works reasonably well; in the case of a sharp discontinuity, however, the method breaks down.
\cite[p. 1398]{d89}.

The sudden sense of “radical novelty” \cite{d89} forms an unexpected challenge for many learners, presenting a barrier to learning. This is because those without prior experience need to adapt to thinking about the intangible and abstract concepts which are needed to describe the mechanics behind the code they are writing \cite{db89}. Barriers can even arise as early as the first stage of instruction. Consider how someone new to reading program code might conceive the mechanics behind an assignment operation, such as: \\[0.5pt]

a = 1;

b = 2;

a = b;

What is the value of a? \\[0.5pt]

Bornat, Dehnadi and Simon found that for ``simple'' assignment operations that ``hardly look as if they should be hurdles at al'', students held many different mental models for how the program may execute \cite[p. 54]{bds10}. Even after a few weeks of instruction, some participants failed to apply the correct model consistently in a diagnostic test. This illustrates that the ways in which learners conceptualise computer programs can be diverse and incorrect models may persist without some intervention. Consequently, it is important not to dismiss the early challenges experienced by individuals as: trivial; a lack of effort; or a lack of talent. Put elegantly, ``if students struggle to learn something, it follows that this is for some reason difficult to learn'' \cite[p. 53]{j02}. These issues can be addressed through soft scaffolding, such that individual understandings are continuously probed to enable the timely delivery of tailored support \cite{sk07}. Through this, misunderstandings are traced and corrected through the provision of intermediate learning objectives. When not promptly addressed, such issues can impede progress as learners are forced to the edge of, or perhaps beyond, their ``zone of proximal development'' \cite[p. 86]{v78}.

Yet, Kinnunen and Malmi note there can be ``individual variety in how students respond to the same situation'' \cite[p. 107]{km06}. Many learners who encounter such challenges are able to overcome them without assistance, albeit perhaps after some frustration. So why are some people tenacious while others seem helpless? A potential candidate for mediating this response is an individual's academic beliefs. Notably, implicit beliefs surrounding programming aptitude. Dweck \cite{d02} divides learners into entity-theorists, who believe their aptitude is a natural fixed trait, and incremental-theorists, who believe their aptitude is a malleable quality which is increased through effort. These two groups demonstrate different behaviours when they encounter difficulty \cite{d02}, as summarised in Table \ref{tab:mindsets}:

\begin{table}
\caption {Potential Influences of Diffent Mindsets (Adapted from \cite{d02})} \label{tab:mindsets} 
\small
\centering
\begin{tabular}{ L{11em} L{12em} L{12em}} 
\hline
	 					& \textbf{Entity-Theorists}				& \textbf{Incremental-Theorists} \\
\hline
  Goal of the Student?			& To demonstrate high coding ability 			& To improve coding ability, even if reveals poor progress \\
  Meaning of Failure? 			& Indicator of low programming aptitude		& Indicative of lack of effort, poor strategy, or missing pre-requisite \\
  Meaning of Effort? 			& Demonstrates low programming aptitude		& Method of enhancing programming aptitude \\
  Strategy when meets difficulty?		& Less time practicing					& More time practicing \\
  Performance after difficulty?		& Impaired							& Equal or improved \\
\hline
\end{tabular}
\end{table}

Too often, it is the case that learners start to believe an inherent aptitude is required to become a programmer. Such beliefs inhibit practice. Thus, it is important that programming pedagogies reinforce the incremental theory. An example might include the liberal use of detailed informative feedback. This approach focuses on improvement through illustrating weaknesses to overcome, rather than merely labeling learners with summative grades. The latter might be interpreted as a judgment of aptitude. However, many learners ``often focus on topics associated with assessment and nothing else'' \cite[p. 14]{gs04} so some form of marking is often necessary as an extrinsic motivator.

While Dweck's \cite{d02} dichotomy is useful in illustrating some differences, it does not explain why some learners seem far more determined than others. Potential factors are the negative affective states that learners can experience as they write code \cite{h04,rs10}. These ``states such as frustration and anxiety [can] impede progress toward learning goals'' \cite[p. 698]{mll07}. However, while some learners become overtly frustrated with the all or nothing nature of preparing a computer program for compilation, others press on without complaint, demonstrating an admirable level of experimentation and debugging proficiency. This can be somewhat surprising given that anything short of a completely syntactically correct set of coded instructions will result in failure and it is unusual for those at an introductory level to write robust code on their first attempt.

A potential candidate for mediating how learners are able to overcome negative affect is academic self-concept. That is, ``self-perceptions formed through experience with and interpretations of one's environment'' \cite[p. 60]{mm11}. Many domain-specific forms of self-concept demonstrate a reciprocal relationship with academic achievement in their respective areas \cite{mm11} as well as, more generally, interactions with study-related emotions \cite{g+10}. Extending this notion, learners who believe that they are programmers, those with a high programming self-concept, may be able to overcome frustrations and anxiety more easily. Thus, maintaining high levels of motivation. However, how can self-concept be enhanced? A meta-analysis of 200 interventions shows that practices which target a domain-specific facet of self-concept, with an emphasis on motivational praise and feedback alongside skill development, yield the largest effects \cite{o+06}. Other aspects of effective practice might also emphasize learning activities that are enjoyable and nurture senses of pride \cite{g+10}.

\section{Conclusion}

Learners often need to practice writing code frequently in order to acquire basic programming competencies. This paper questions how learning environments can be better designed in order to facilitate deliberate practice, describing three potential barriers to deliberate practice: the radical novelty of the learning material; the belief that some inherent aptitude is required; and the emergence of unfavourable affective states. Overcoming such barriers will facilitate educators in aiming for excellence, but often require strategies that are analytical and adaptable. It is proposed that examples of good practice include: soft scaffolding; ongoing detailed informative feedback; and an emphasis on self-enhancement, through motivational feedback and pride-worthy activities, in addition to skill-development. 

\bibliographystyle{splncs03}
\bibliography{scott-educatingprogrammers}

\end{document}